\documentclass[11pt,twocolumn]{article}
\usepackage{pslatex, graphicx, amssymb, amsmath}
\usepackage{enumitem}
\usepackage{epsfig}
\usepackage{cite}
\usepackage{xcolor}
\usepackage {amsmath}
\usepackage{caption}  
\usepackage{multirow}
\usepackage{amsfonts}
\usepackage{subcaption}
\usepackage{color}

\usepackage{footnote}
\usepackage{algorithm}
\usepackage{algorithmicx}
\usepackage{booktabs}
\usepackage{booktabs}
\usepackage{booktabs}
 \usepackage{multicol}
\newcommand\Tstrut{\rule{0pt}{2.4ex}}         
\newcommand\Bstrut{\rule[-0.9ex]{0pt}{0pt}}
\def\abstract{
\typeout{Abstract}
 {\bf Abstract} 
} 

\begin{document}
\title{Exploiting ML algorithms for Efficient Detection and Prevention of JavaScript-XSS Attacks in Android Based Hybrid Applications}

\author{Usama Khalid \\ Department of Computer Science, COMSATS University Islamabad, Wah Campus, Pakistan \and Muhammad Abdullah\\ Department of Computer Science, University of Engineering and Technology Taxila, Pakistan 
        \and Kashif Inayat\\ Department of Electronics Engineering, Incheon National University, South Korea\thanks{ 
        	Corresponding author: Kashif Inayat (email: kashif.inayat@inu.ac.kr).}}

\maketitle
\begin{abstract}
The development and analysis of mobile applications in term of security have become an active research area from many years as many apps are vulnerable to different attacks. Especially the concept of hybrid applications has emerged in the last three years where applications are developed in both native and web languages because the use of web languages raises certain security risks in hybrid mobile applications as it creates possible channels where malicious code can be injected inside the application. WebView is 
an important component in hybrid mobile applications which used to implements a sandbox mechanism to protect the local resources of smartphone devices from un-authorized access of JavaScript. However, the WebView application program interfaces (APIs) also have security issues. For example, an  attacker can attack the hybrid application via JavaScript code by bypassing the sandbox security through accessing the public methods of the applications. Cross-site scripting (XSS) is one of the most popular malicious code injection technique for accessing the public methods of the application through JavaScript. This research proposes a framework for detection and prevention of XSS attacks in hybrid applications using state-of-the-art machine learning (ML) algorithms. The detection of the attacks have been perform by exploiting the registered Java object features. The dataset and the sample hybrid applications have been developed using the android studio. Then the widely used toolkit, RapidMiner, has been used for empirical analysis.  The results reveal that the ensemble based Random Forest algorithm outperforms other algorithms and achieves both the accuracy and F-measures as high as of 99\%.\end{abstract}

\section{Introduction}\label{sec-1}
%
%
%
%
In the past few years, the rate of using smart phones and android applications has increased tremendously. Different new techniques in development of android applications, have been used, some of them make smart pones vulnerable for different attacks (especially XSS attacks). There are two main types of XSS attacks: persistent XSS attack and non-persistent attack  \cite{b1}. In persistent attacks, an attacker stores malicious script in a server and when a victim accesses that server, the malicious script starts working in the victim's browser and important information is sent to the attacker in return. In non-persistent attacks, an attacker sends malicious script to a victim's browser, and when the victim runs that script, the web application opens, where the attacker steals important information. According to the report of WhiteHat Security \cite{b2}, almost 432 million Android smartphones have been sold in year 2017. This study also reports that almost 30\% attacks are involved in web applications (asynchronous JavaScript and extensible markup language (AJAX) intrusion \cite{b3}, phishing attacks etc.), 62\% attacks are based on feature hacking to exploit vulnerabilities, 81\% attacks are hacking related breaches that leveraged weak or stolen passwords and 32\% attacks are exploiting structured query language (SQL) injection errors in dynamic web applications \cite{b2}. Another report by Gartner indicates that more than 50\% mobile applications are using hybrid technology  \cite{b4}. Hybrid applications are developed using both Java and web languages such as hypertext markup language (HTML), cascading style sheets (CSS) and JavaScript  \cite{b5,b6}.
The structure of hybrid applications is similar to web applications. However, the main component in hybrid applications is WebView  \cite{b4,b7,b8,b9}, which is provided by the browser engine named WebKit. WebView provides the basic browser functionality to load and display web pages within Android applications without switching to the default browser. This WeView based facility decreases the loading time of web pages. More importantly, Android applications can interact with JavaScript code embedded \color{black} in web pages directly  \cite{b4}. WebView is placed inside a native container. Therefore, it can easily access mobile hardware resources such as mobile libraries, messages, contacts with the help of APIs. Since WebView uses different APIs to access resources and there exist potential loopholes in these APIs, most hackers use XSS attacks to hack different types of smart phone devices  \cite{b2,b10}.
Although the XSS attack mechanisms in hybrid applications are a little bit different from those in normal web attacks, the main concept behind the attacks is the same. Most common attacks in hybrid applications \color{black} are invoking native Java from JavaScript (through WebView) and invoking JavaScript from native Java (through WebView)  \cite{b9}. In hybrid applications while invoking Java from JavaScript, an attacker stores the malicious JavaScript code in an web page, when the web page opens in the application and then an object is registered in the WebView. First of all, the attacker accesses that object through JavaScript and through that object the attacker accesses the native method of that application, from where the attacker can easily get different smartphone resources by triggering important built-in functions inside the native method.\\
\subsection{General Idea}
To address the aforementioned research issues, we propose an XSS detection and prevention framework. The steps of the proposed approach are as follows. First of all, when a Java object is registered in WebView, threat prevention unit in our prevention system extracts those features of the object and sends these feature information to the detection unit. Then the detection unit applies machine learning based algorithm on these features and detects whether the incoming object is accessing important native methods or not, and sends the resulting decision to the threat prevention unit. In the case of detection of the attack, the threat prevention unit becomes active and the alert application asks the user if this object is allowed to get registered and access the application or not. If the user response is yes, then the threat prevention unit allows this object to access phone resources. Otherwise, the threat prevention unit blocks that object and records it as an attack for future attack detection.
\subsection{Contributions}
The major contributions of this paper can be
summarized as follows:
\begin{itemize}
	\item We generated dataset by analyzing records behavior of 500 different applications and then extracted 7 different features from the dataset.	
	\item To detect all threats that come from web pages, we hooked the addJavascriptInterface API and detection framework. Therefore, every Java object that is intended to be registered into WebView can be inspected.\color{black}
	\item We used 7 ML algorithms (Evolutionary-Support Vector Machine (E-SVM), Neural Network, Naive Bayes, Support Vector Machine (SVM), Bagging, Random Forest and J48) for the malicious APIs classification in the detection unit, algorithms learn and efficiency increase with time.
	\item We made the analysis on all algorithms based on accuracy, F-measure and execution time and conclude the Random Forest is the best among detecting XSS attacks. 	
\end{itemize}
\subsection {Organization}\label{sec-1.2}
The organization of this paper is as follows: Section \ref{sec-2} reviews related work.  Section \ref{sec-3} describes the preliminaries, which includes sandbox mechanism, the WebView component and describes how WebView bypasses the sandbox mechanism. Section \ref{sec-4} propose our framework which consists of detection and prevention of XSS. Tools in use and evaluation dataset are explained in Section \ref{sec-5}. Our analysis results are presented in Section \ref{sec-6}. Section \ref{sec-7} describes which algorithm is better for our detection system. 
concludes this paper.\color{black}
\section{Related Work}\label{sec-2}
The relevant literature presents a number of research studies to detect and prevent the security attacks on the hybrid applications. We have differentiate several examined approaches into three categories: a) classification based techniques b)
simple techniques, and c) soft computing based techniques. 
\subsection{Classification based techniques}\label{sec-2.1}
There are a lot of mobile frameworks which are used in development. Most android hybrid applications are developed with the PhoneGap framework, through which attackers can easily bypass the sandbox mechanism. That is why hybrid applications provide a way to bypass access control policies of both WebView and WebKit, and run malicious code in user’s application. In the PhoneGap plugins, malicious code can access and steal a user’s private information and destroy the user’s file system. In hybrid application, JavaScript is also used in development. Therefore, attackers can launch their attack through encoded JavaScript in human-unreadable form on rendering web pages in user’s application. 
On classifying different types of applications result came up with 95.3\% precision. That means most of the applications are suffering from this issue \cite{b4, b11}. As different type of web sites render in WebView in hybrid applications, attackers can also launch attack through malicious URLs and then by applying improved semisupervised algorithm to construct URL multi-classification model. Hybrid applications can detect and blacklist
URLs through efficient URL classification in \cite{b12}. Different web sites rendered in WebView consist of HTML tags, script functions, hyperlinks and advanced features. However, these features increase the security risk. Features extracted by Angelo et al. in [13] improve the accuracy of automatic XSS classification by 98.5\%.
\subsection{Simple techniques}\label{sec-2.2}
WebView is an essential component in hybrid applications which provide different APIs through which application can interact with web pages. This interaction allows to access resources using Java Object. Access control on security-sensitive APIs on the Java object level uses static analysis to detect these sensitive APIs at runtime and notifies a user if it detects any threat \cite{b7, b9}. Attackers can launch attacks using different strategies, one of which is SQL injection. 
Inyong et al. removes the value of an SQL query attribute of web pages when parameters are submitted and then compares it with the predetermined one. This method shows an effective results \cite{b17, b19}. SQL injections can also be prevented by using a technique called Sania during development and debugging. It intercepts SQL query between application and database, and generates attacks according to the syntax of vulnerable spots in SQL. It also compares parse trees of SQL with parse trees after attacks to evaluate the safety of those spots. Another way to block SQL injection is by applying AMNESIA, which is based on combination of static and dynamic analysis. In static phase, model of all query strings is extracted from web application code using existing string analysis. In dynamic phase, it monitors dynamically-created queries for conformance with statically-build model. Quires which are not complaint are identified as SQL Injection Attacks (SQLIA's) \cite{a17, c17}. Another emerged strategy is the ADSandbox system which detects malicious web sites attacks through JavaScript by logging every critical action of web site. Using these logs, ADSandbox decides whether the web site is malicious or not \cite{b14}. Android application (AA) sandbox is also an efficient strategy which performs both static and dynamic testing in Android programs to automatically detect suspicious applications \cite{b15}. Different hybrid applications are developed in multiple languages with different semantics which may be vulnerable. HybriDroid, a static analysis framework, investigates semantics especially for the interoperation mechanism of Android Java and JavaScript. This framework analyzes inter-communication between Android Java and JavaScript. \cite{b8} Another framework called Crowdroid is for collection of traces. It works on two types of datasets: artificial malware created for test purposes and real malware found in the wild. In \cite{b20}, a simple technique has been proposed to detect the XSS attack in an efficient way. Another framework called Cordonove combines HTML5, JavaScript (JS) and native application code to develop hybrid application. However, combining languages increases security threats. 
Achim et al. method constructs a uniform call graph for hybrid applications through which the system can detect malicious calls \cite{b22}. On the other hand, there is an insufficiency of access control in HTML5 applications. To overcome this insufficiency, the fine-grained access control mechanism separate subjects within the same application by defining frame-based and origin-based policy \cite{a22}. There are a lot of hidden injection points in HTML5 based web applications through which attackers can attack and steal important information. Context-Sensitive sanitization based XSS framework can detect those injection points, sanitize them and remove XSS attacks \cite{b30}. With the exploding number of android applications attackers can launch XSS attack through malware calls. Therefore It’s very important to analyze and detect malicious behaviors of applications which launch these calls for analysts. Mobile-Sandbox covers this whole perspective in two ways: by combining static and dynamic analysis; and by using specific techniques to log these malicious calls which trigger native API's. We run 36,000 applications through this system and results came up with 24\% applications use native calls in \cite{b16}. In hybrid applications, permissions are not enough to ensure the security of private information. 
SCANDAL is a sound automatic static analyzer which detects privacy leaks in Android applications and alerts users if threats are found. It analyzed real-time applications and detected privacy leaks in 11 applications and 8 known malicious applications
\cite{b21}. Malicious applications can be third party applications which can access and steal sensitive data. TaintDroid provides real-time analysis of third-party applications with 68 instances of potential misuse detected \cite{b28}.
\subsection{Soft computing based techniques}\label{sec-2.3}
Artificial intelligence plays a vital role in detection and prevention of XSS attacks in hybrid applications. 
A search based approach for security testing of web applications by using static analysis to detect XSS attack is presented in \cite{b24}. Genetic Algorithm (GA) pass inputs automatically which expose those vulnerabilities. By using search results test cases are presented to developers to understand and fix these issues.In \cite{b25}, another approach to expose XSS attacks using genetic algorithm is by creating XSS attack patterns is proposed. By using test data, if it uncovers any path, then it records as malicious. Therefore, genetic algorithm-based test data generator uses a database of XSS attack patterns to generate possible attacks and assesses whether the attack is successful or not. In \cite{b26}, classification of Android’s application files have been done through machine learning techniques, static features are extracted from Android’s Java byte-code and other file types such as XML-files. They performed evaluation on 2850 applications and gained an accuracy level of 91.8\%. One of the popular machine learning technique is data mining which can also detect XSS attacks. The combination of data mining and taint analysis techniques to detect false positives generated 5\% better results than PhpMinerII's and 45\% better results than Pixy's \cite{b27}. We can also improve our XSS
detection results by combining genetic algorithm and static analysis. Pixy is used to detect vulnerabilities and uncover existing XSS attack by removing infeasible paths and using test data. Then genetic algorithm is applied to only those
paths to detect XSS Attacks \cite{b31}. Finally XSS attacks can also be detected by combining model of inference and evolutionary fuzzing. Inference model is used to obtain knowledge. Based on this knowledge, a genetic algorithm generate
inputs with better fitness values \cite{b32}.\\\\
\section{Preliminaries}\label{sec-3}
To understand that how XSS attacks are launched through WebView component, first of all we
have to understand the following two main concepts: sandbox mechanism (WebView security) and WebView component (working of WebView).
\subsection{WebView}\label{sec-3.1}
The WebView is mainly powered by a browser engine which is named as WebKit. WebView provides a basic browser functionality to load and display web pages within Android applications without switching to the default browser. In simple words, Whenever a user clicks on any type of link in the application, in convention methods the application closes and opens that link in a mobile browser. Application takes time  this to perform this whole process. To remove this time delay, Android companies introduced a built -in mini browser called WebView. It is a custom and powerful browser inside the application. The main thing to notice here is, Android application can interplay with JavaScript code embedded in web pages by using the WebView APIs. APIs are basically used in the interaction between the Android application and web pages. Those APIs include setJavaScriptEnabled API, addJavascriptInterface API, and loadUrl API  \cite{b7}. Where, SetJavaScriptEnabled API enables the JavaScript, addJavascriptInterface API registers the Java object, and loadURL API loads web pages.
\subsection{Sandbox}\label{sec-3.2}
Sandbox is a tool which manage the access control of the web browsers (e.g., WebVew or Android browser). It isolates the web pages which are running inside the WebView, \cite{b4,b7,b9,b14}. 
\begin{figure}[ht]
	\centering{\includegraphics[width=\columnwidth]{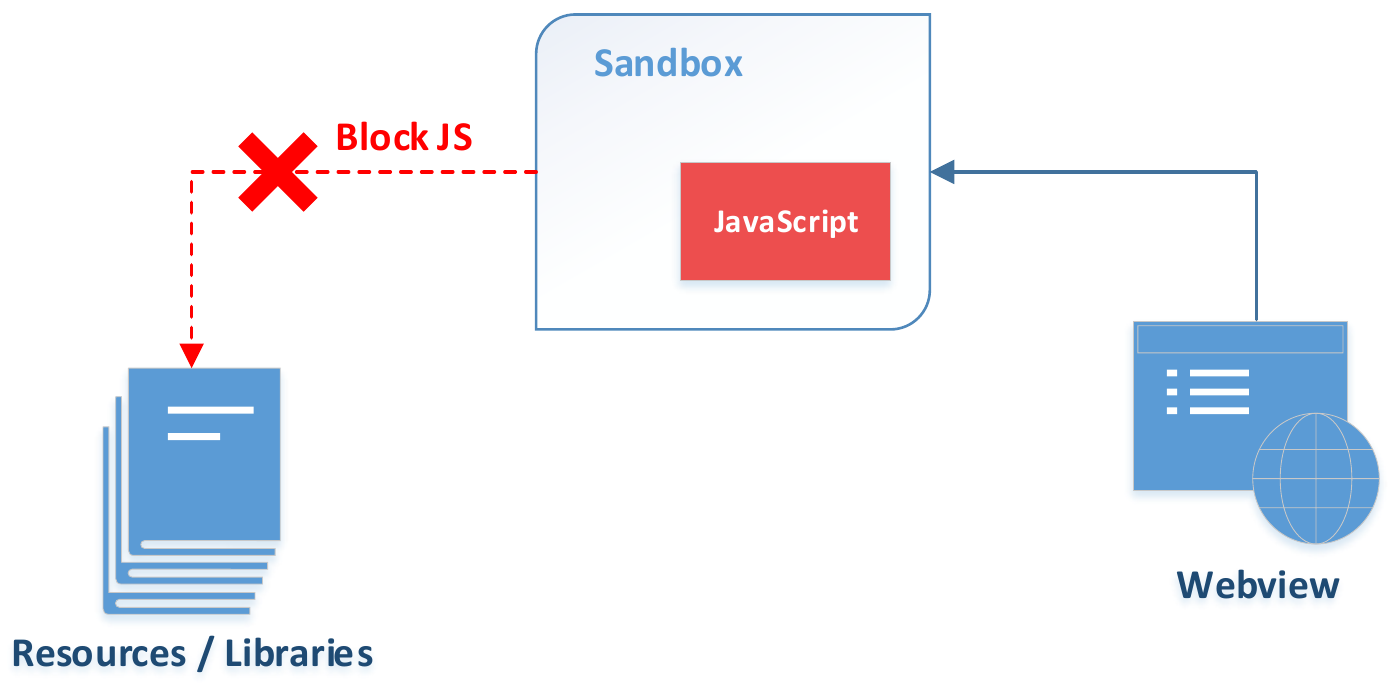}}
	\caption{\textcolor{black}{Sandbox mechanism}.}
	\label{fig1}
\end{figure}
Therefore, if there is a malicious JavaScript running inside WebView through a web page, sandbox will isolate that JavaScript and block it to invoke the access to the mobile local resources, as shown in Figure \ref{fig1}  . Then sandbox mainly used for the following two objectives:
\begin{itemize}
	\item Isolate web pages from the system and isolate web pages of one origin from those of another.
	\item Enforce the same-origin-policy (SOP). 
\end{itemize}
\subsection{Bypassing sandbox mechanism}\label{sec-3.3}
The WebView provide fast browsing services Android applications but opens Pandora box of security vulnerability. 
\begin{figure}[ht]
	\centering{\includegraphics[width=\columnwidth]{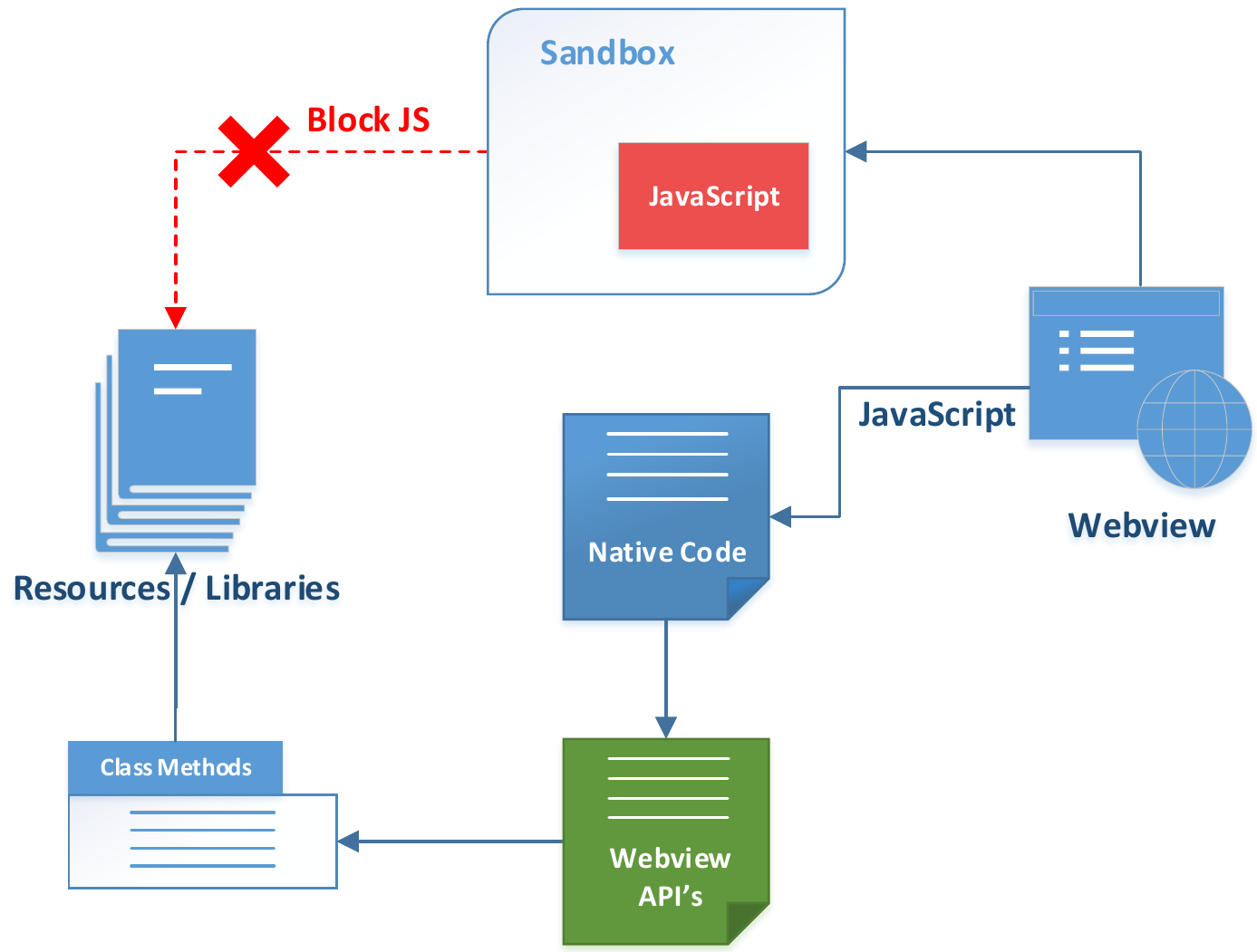}}
	\caption{\textcolor{black}{An illustration of the bypassing sandbox mechanism}.}
	\label{fig2}
\end{figure}
In Figure \ref{fig2}, it can be seen that despite the presence of sandbox security mechanism in WebView to protects the local resources of mobile from JavaScript attacks, WebView itself bypasses the sandbox mechanism with the help of WebView afformentioned APIs. This creates a path through which JavaScript can easily communicate with native Java. And an attacker can trigger different class methods of Java to launch XSS attack to access the important functions of mobile including local resources trough native Java code. In simple words, the attacker can easily launch this kind of attack by running malicious web site inside WebView with the help of malicious JavaScript.
\section{The Proposed Framework}\label{sec-4}
The proposed framework is organized as follows: 1) Steps to detect attacks on sensitive API's.
2) Detection framework which detects attacks on API's. 3) Prevention framework which will prevent an attack.
\subsection{Steps and Challenges}\label{sec-4.1}
As discussed in the Section \ref{sec-3}, in WebView main threat is due to use of different types of WebView APIs. The most vulnerable and important APIs is addJavaScriptInterface (), which registers the Java object. Therefor, we proposed the control of thosse Java object which is necessary to prevent such vulnerabilities, which lead us the following contrivances:
\begin{enumerate}
	\item Need to impose a framework at the Java object level to confine java object activities.
	\item	Need to detect if Java object is accessing important API which can cause harm to the user information.
\end{enumerate}
To perform the above two control schemes, we had to understand and address the following terms and challenges:
\begin{enumerate}[label={\alph*)}]
	\item We need to understand that how WebView APIs can be executed by Java objects?
	\item What type of vulnerable APIs can be triggered through Java objects?
	\item	How to detect and What to do if a threat is detected at Java object level?
\end{enumerate}
To address the challenge (a), we need to understand Java object registration, JavaScript, enabling and loading of the website. By default JavaScript is disabled in WebView, the API named as addJavascriptInterface() helps the Java object to be registered in WebView. To make JavaScript enable, APIsetJavascriptEnabled() is set to true. Then API LoadURL() loads the web site inside WebView container. Once, a JavaScript object is registered with loaded URL and JavaScript is enabled, JavaScript inside the web site can easily access the registered object which subsequently can access native methods from where JavaScript will trigger/execute WebView  APIs \cite{b9}.

To address the challenge (b), we have to identify all those APIs through which an attacker can steal information. Most vulnerable APIs have been enlisted in Table \ref{tabl1}.
\begin{table}[ht]
	\centering\scriptsize
	\caption{A list of vulnerable APIs which provide access to the sensitive information.}\label{tabl1}
	\smallskip
	\begin{tabular}{|c|c|}
		\hline
		
		\multicolumn{2}{|c|}{Vulnerable API's}\Tstrut\Bstrut\\\hline\hline		
		getCellLocation&getAccounts\Tstrut\Bstrut\\
		getDeviceId&getAuthToken\\
		getPhoneType&getUserDate\\
		getSubscriberId&peekAuthToken\\
		getLine1Number&removeAccount\\
		getSimSerialNumber&setPassword\\
		getVoiceMailAlphaTag&getName\\
		getVoiceMailNumber&getProfileConnectionState\\
		SendTextMessage&getProfileProxy\\
		sendMultipleTextMessage&getParams\\
		sendDataMessage&getUnzippedContent\\
		getAllProvider&getCertificate\\
		getBestProvider&clearHistory\\
		getGpsStatus&clearSearches\\
		getLastKnownLocation&getAllBookMarks\\
		clearPassword&getAllVisitedUrls\\
		editProperties&getNetworkOperator\\
		semdMultipartTextMessage&\\\hline
		
	\end{tabular}
\end{table}
To address the challenge (c), we proposed the feature extraction and ML techniques have been considered to use to detect these types of attacks through extracted features. Main concept of threat detection is same as explained in  \cite{b7}\footnote{The only difference is authors in this article used statistical model for the detection.} at runtime to detect whether threats exist in Java object or not and the permission request to the user accordingly. As ML algorithm used for the classification which will detect the attack in detection unit and sends the result to prevention framework. Then prevention framework will ask the user to allow it or not. If the user gives a permission, it will go on or block that attack.
\subsection{Detection Framework}\label{sec-4.2}
Detection framework is divided into 4 levels. In level 1, collection of dataset and pre-processing is performed and saved in the database. Then in level 2, we apply
\begin{figure}[ht]
	\centering{\includegraphics[width=\columnwidth]{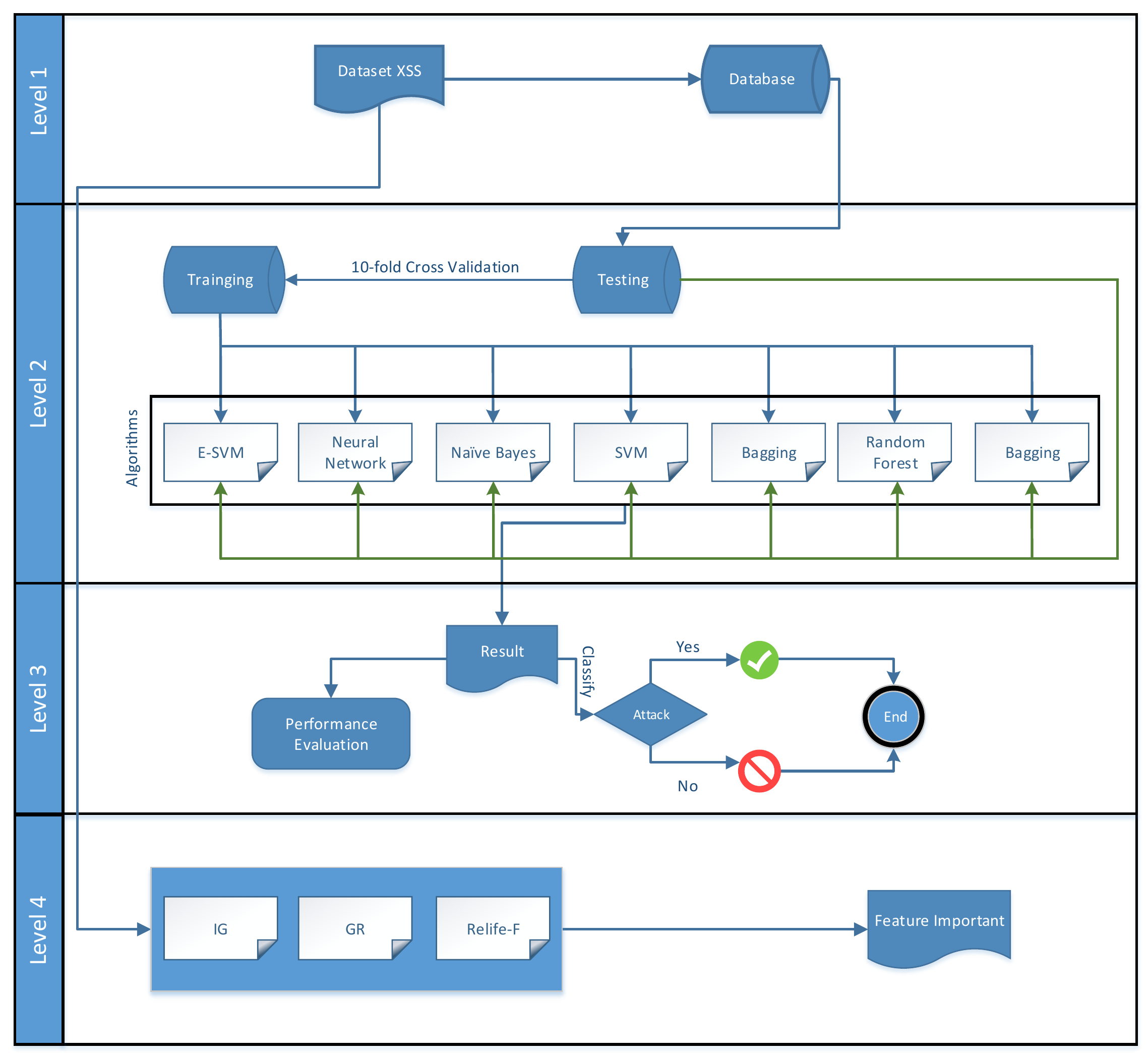}}
	\caption{\textcolor{black}{An architecture of the detection framework for attack detection}.}
	\label{fig3}
\end{figure}
10-fold cross validation using ML algorithms. We use different 7 ML algorithms (Evolutionary-Support Vector Machine (E-SVM), Neural Network\footnote{Artificial Neural Network (ANN)}, Naive Bayes, Support Vector Machine (SVM), Bagging, Random Forest, and a classical probability based algorithm, J48) to perform the classification. The main reason to use ML algorithms is that after providing samples, algorithms learn and detection efficiency improve accordingly. Moreover, based on different performance measures, best algorithm have been identified. Performance evaluation have been performed in third level 3 and also at the same time provided the classification result to  prevention framework. In level 4, selection algorithms (Information Gain (IG), Gain Ratio (GR) and Relief-F (RF)) have been used to identify the importance of features. This help to identify which features play an important role in detecting the XSS attacks.
\subsection{Prevention Framework}\label{sec-4.3}
\begin{figure}[ht]
	\centering{\includegraphics[width=\columnwidth]{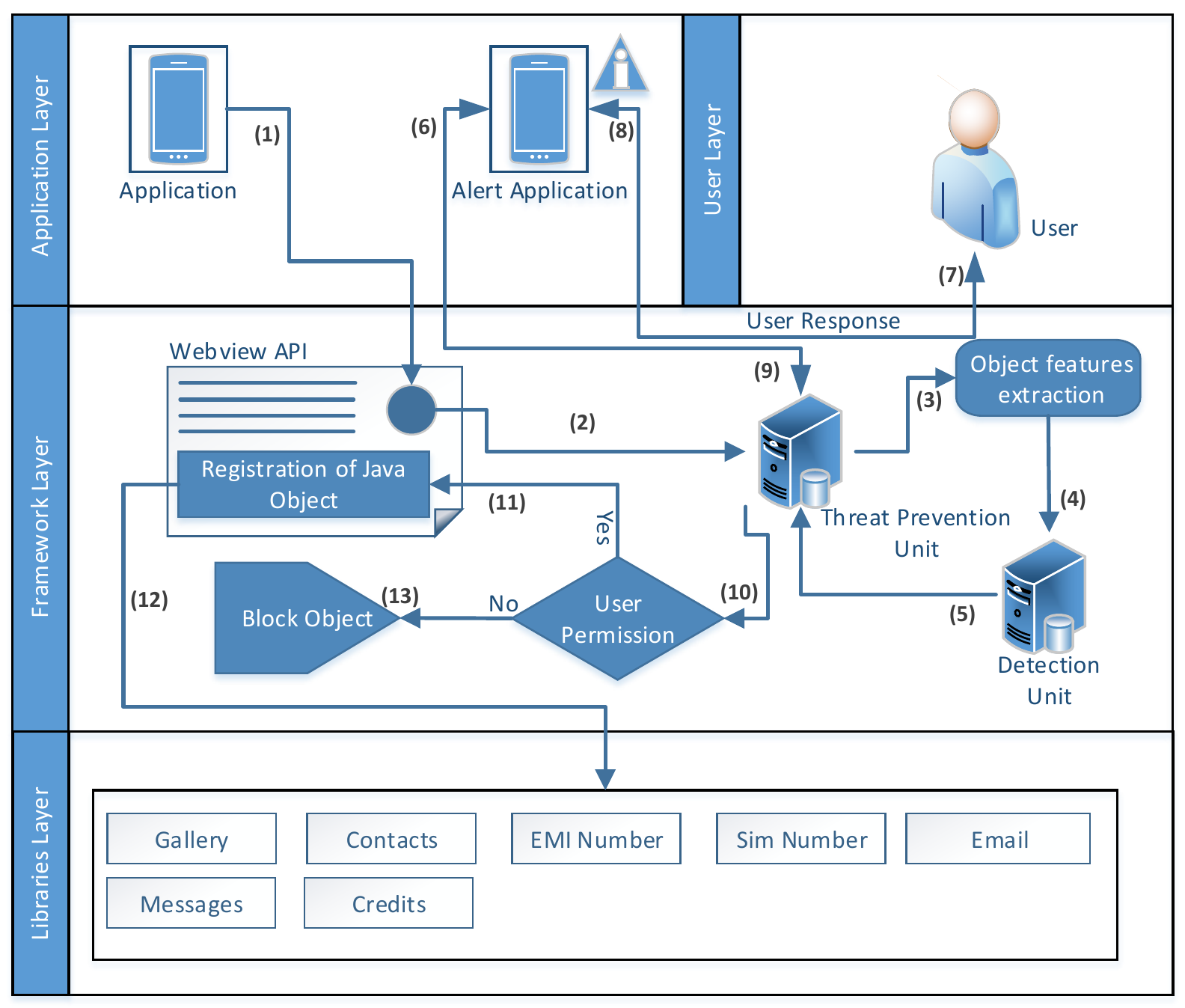}}
	\caption{\textcolor{black}{An architecture of the prevention framework for attack prevention}.}
	\label{fig4}
\end{figure}

Prevention framework is shown in Figure \ref{fig4}. It consist of three parts: applications layer, framework and libraries. The following steps explain the whole process and  flow of prevention framework:
\begin{enumerate}
	\item At first Android application open a malicious web site inside WebView, and call the addJavaScriptInterface() API. 
	\item Before completing registration, information regarding Java object go to threat prevention unit.
	\item 	Then features extractor extracts the features of the Java object from Threat Prevention Unit and provide features information to detection unit. 
	\item Detection Unit gets all of the features and applies classification on that features using ML algorithm, as explained earlier. After that it sends final result (attack or not [Yes / No]) to Threat Prevention Unit, so in this way detection unit intercepts the call to the addJavaScriptInterface() API.
	\item 	If Threat Prevention Unit receives 'Yes' from Detection Unit, the proposed method calls the alert application: otherwise it proceeds to step (9), to allow automatically and then proceeds to step (10) to complete the Jave object registration.
	\item Alert Application warns the user about the threat, which shows information about attacking web page Name, Object Name, Security Sensitive API calls by the object.
	\item The user replies to Alert Application to decide whether to disable the Java Object or not.
	\item Alert Application forwards the user's decision to Threat Prevention Unit.
	\item 
	Threat Prevention Unit proceed further whether to disable the object or not on the basis of the users' decision.
	\item If Threat Prevention Unit's decision is 'Yes', then it allows the object to register and object get access methods of Java class from which it can access different mobile libraries.
	\item If Threat Prevention Unit takes a decision 'No', then it disables the Java object and block that web site. 
\end{enumerate}

\section{Experimental Setup }\label{sec-5}
In the whole experiments, we used different tools, dataset and machine learning algorithms (we omit the details about ML algorithms to keep the main focus on detection and prevention). Evaluation of ML algorithms also have been performed in subsequent section.
\subsection{Tool}\label{sec-5.1}
We used the mainly RapidMiner software platform, it unites data preparation, machine learning, and predictive model deployment. Organizations can build machine learning models and put them into production faster than ever.Rapid Miner’s also lightning fast visual work-flow designer and have automated modeling capabilities.
\subsection{Dataset}\label{sec-5.2}
We created our own dataset named as APK\_XSS\_ATTACK, where APK is Android Package Kit. To collect benchmark data, demo victim application have been created and launched a live attack on it, and then recorded those live attacks at run time. We took 17 live attacks entries and recorded the patterns for XSS attacks, in this way we generate 444 raw XSS attacks to meet standard dataset requirements. In this dataset, the attack ratio is 50\%. We have gathered 460 samples on 20 APK’s.
\begin{table}[ht]
	\centering\scriptsize
	\caption{Characteristics of attacks.}\label{tabl2}
	\smallskip
	\begin{tabular}{|c|c|}
		\hline
		\Tstrut
		Categories &Total \\\hline\hline
		\Tstrut
		Total XSS Attacks& 460 \\
		XSS Attacks & 230 \\
		Non XSS Attacks & 230 \\\hline
		
	\end{tabular}
\end{table}
As Table \ref{tabl2} shows  the characteristics of dataset, we recorded total 460 attacks. From these, we recorded total 230 XSS attacks (accessing sensitive APIs) and 230 non-XSS attacks (not accessing sensitive APIs).
\begin{table}[ht]
	\centering\scriptsize
	\caption{List of features of dataset.}\label{tabl3}
	\smallskip
	\begin{tabular}{|c|p{5cm}|}
		\hline\Tstrut
		Feature Name & Description \\\hline\hline \Tstrut
		App Names & Name of victim applications \\\hline\Tstrut
		Permissions & Include all webview permissions \\\hline\Tstrut
		API Name & Name of all sensitive API. \\\hline\Tstrut
		Website Name & Name of malicious websites. \\\hline\Tstrut
		IP & IP of website. \\\hline\Tstrut
		\multirow{2}{*}{Location}& Country Location of website. (From which country attack is launched) \\\hline\Tstrut
		Label & Class of attack ( Yes / No) \\\hline
	\end{tabular}
\end{table}
In total, 6 main features have been extracted which are described in Table \ref{tabl3}. One of the most important feature is API Name, which records the name of that sensitive API which gives access to sensitive information.
\subsection{Dataset Conversion:}\label{sec-5.3}
To apply ML algorithm, we converted our nominal data to numerical form by using component (Nominal to Numerical) in Rapidminer. 
\subsection{Algorithms Evaluation}\label{sec-5.4}
To check how efficiency of ML classifiers we exploited on the XSS training set,  classification accuracy and  F-measure (It further use precision and recall) are taken as the measure parameters. Accuracy, F-measure (with recall and precision) are defined as follows, respectively.
\begin{enumerate}
	\item {\bf Accuracy:} Accuracy is used as a performance measure in the domains of information retrieval and data mining. It depicts the fraction of the results that have been successfully retrieved as
	\begin{equation}\text { Accuracy }=\frac{T P+T N}{T P+T N+F P+F N}.\end{equation}Here, FP, FN, TN, and TP, stand for False Positive, False Negative, True Negative, and True Positive, respectively.
	\item {\bf F-Measure: } F-measure takes precision and accuracy. It may be considered as the weighted average of both values as
	\begin{equation}F=\frac{2 \times \text { precision } \times \text { Recall}}{\text {precision }+\text { Recall}}.\end{equation}
	Where precision and recall explained here:
	\begin{itemize}
		\item {\bf Precision:} Precision is the performance evaluation measure that may be known as the ratio of retrieved documents that are related to the search as
		\begin{equation}\text { Precision }=\frac{T P}{T P+F P}.\end{equation}
		\item {\bf Recall:} Recall, known as sensitivity, is the ratio of related instances that have been retrieved over the total amount of retrieved instances as
		\begin{equation}\text {Recall}=\frac{T P}{T P+F N}.\end{equation}
	\end{itemize}
	
\end{enumerate}

\begin{table*}[ht]
	\centering
	\scriptsize
	\caption{Feature ranking by IG, GR and Relief-F.}\label{tabl4}
	\smallskip
	\begin{tabular}{|c|c|c|c|c|c|}
		\hline
		\multicolumn { 2 } {| c| } { Information Gain } & \multicolumn { 2 } { c| } { Gain Ratio } & \multicolumn { 2 } { c| } { Relief-F }\Tstrut\Bstrut\\\hline
		\Tstrut
		Features & Values &  Features & Values & Featues & Values \\ \hline
		\hline
		API Name & 0.860 & API Name & 0.861 & API Name & 2.875 \Tstrut\Bstrut\\
		Permissions & 0.021 & App Names & 0.108 & Permissions & 0.116 \\
		Location & 0.006 & Permissions & 0.108 & Location & 0.018 \\
		App Names & 0.005 & Website Name & 0.108 & IP & 0.007 \\
		Website Name & 0.005 & IP & 0.108 & Website Name & 0.007 \\
		IP & 0.005 & Location & 0.097 & App Names & 0.006 \\
		\hline
	\end{tabular}
\end{table*}
\section{Results}\label{sec-6}
In this experiment, three dimension reduction techniques are applied to the aforementioned 6 features of the dataset: Information Gain (IG), Gain Ratio (GR), and Relief-F. Based on these tecniques features have been ranked to identify which feature is playing important role in the detection of XSS attacks, ranking have been shown in Table \ref{tabl4}. Moreover, we utilized this information to improve the algorithms accuracy.

API Name is the most important feature as it ranked at first by IG, GR and Relief-F. Permissions feature is ranked at second by IG, third by GR and second by Relief-F. Location feature is ranked at third by IG, 6th by GR and third by Relief-F. App Names is ranked at fourth by IG, second by GR and 6th by Relief-F. Website Name is ranked at 5th by IG, 4th by GR and 5th by Relief-F. Lastly, IP is ranked at 6th by IG, 5th by GR and 4th by Relief-F.
\subsection{Results of classifiers using Cross-Validation}\label{sec-6.1}
As mentioned before, E-SVM, Neural Network, Naive Bayes, SVM, Bagging, Random Forest, and  J48 are applied to the selected features set. 10-fold cross validation is used to train the test dataset. The accuracy and f-measure parameters are used to evaluate the performance of these algorithms. 
\begin{figure}[ht]
	\centering{\includegraphics[width=\columnwidth]{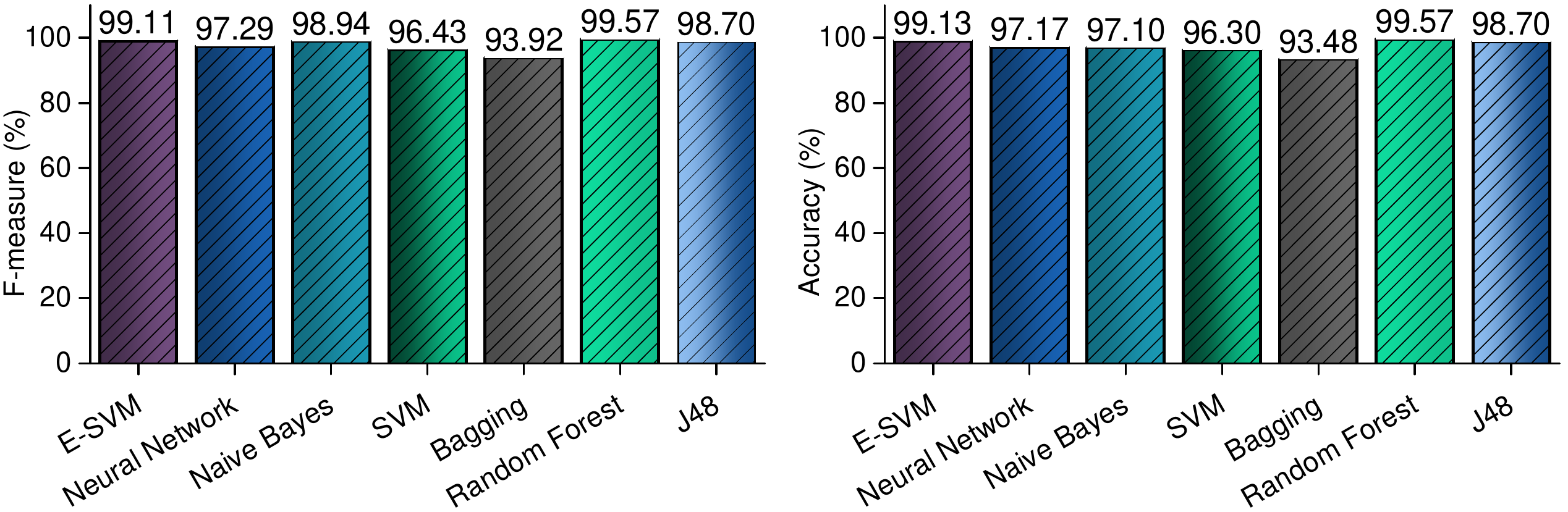}}
	\caption{Accuracy and  F-Measure analysis of ML algorithms.}
	\label{fig5}
\end{figure}
The graph in Figure \ref{fig5} clearly shows that Random Forest performed well in accuracy and f-measure.
Moreover, E-SVM took secod place and others Naïve Bayes and Neural Network also showed good results in the classification process as shown in Figure \ref{fig5}. Therefore, we can conclude that Random Forest is the best algorithm to detects XSS attacks.
\subsection{Comparison of Execution Time}\label{sec-6.2}
\begin{figure}[ht]
	\centering{\includegraphics[width=2.5in]{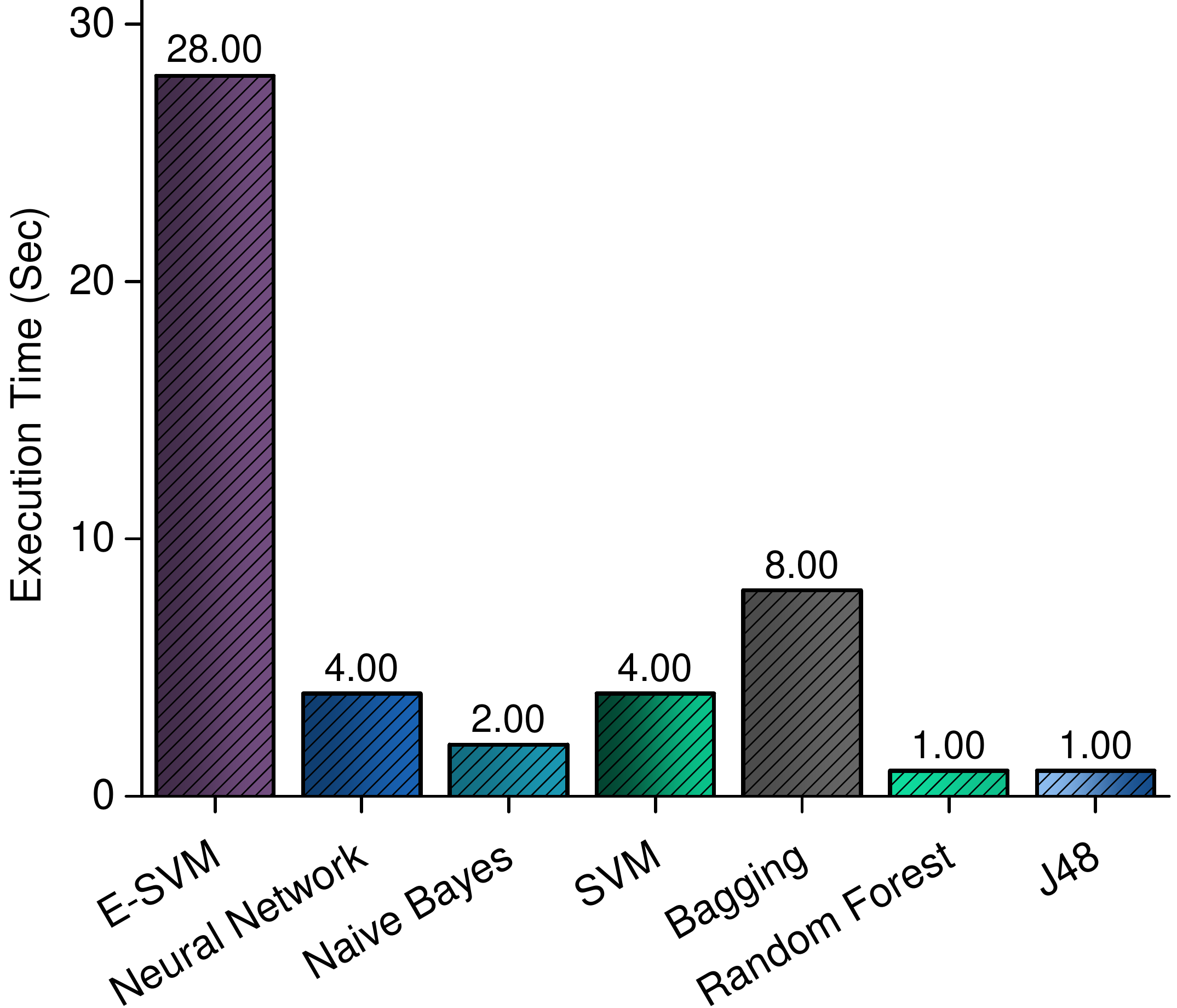}}
	\caption{Execution time analysis of ML algorithms.}
	\label{fig6}
\end{figure}

We also have calculated the average execution time of all classifiers tested on selected dataset shown in Figure \ref{fig6}. The execution time of E-SVM is too high as compared to other algorithms. However, in this graph RF (Random Forest) and J48 lead the all algorithms with just 1 second execution time.
\subsection{ROC based Comparison}\label{sec-6.3}
\begin{figure}[ht]
	\centering{\includegraphics[width=\columnwidth]{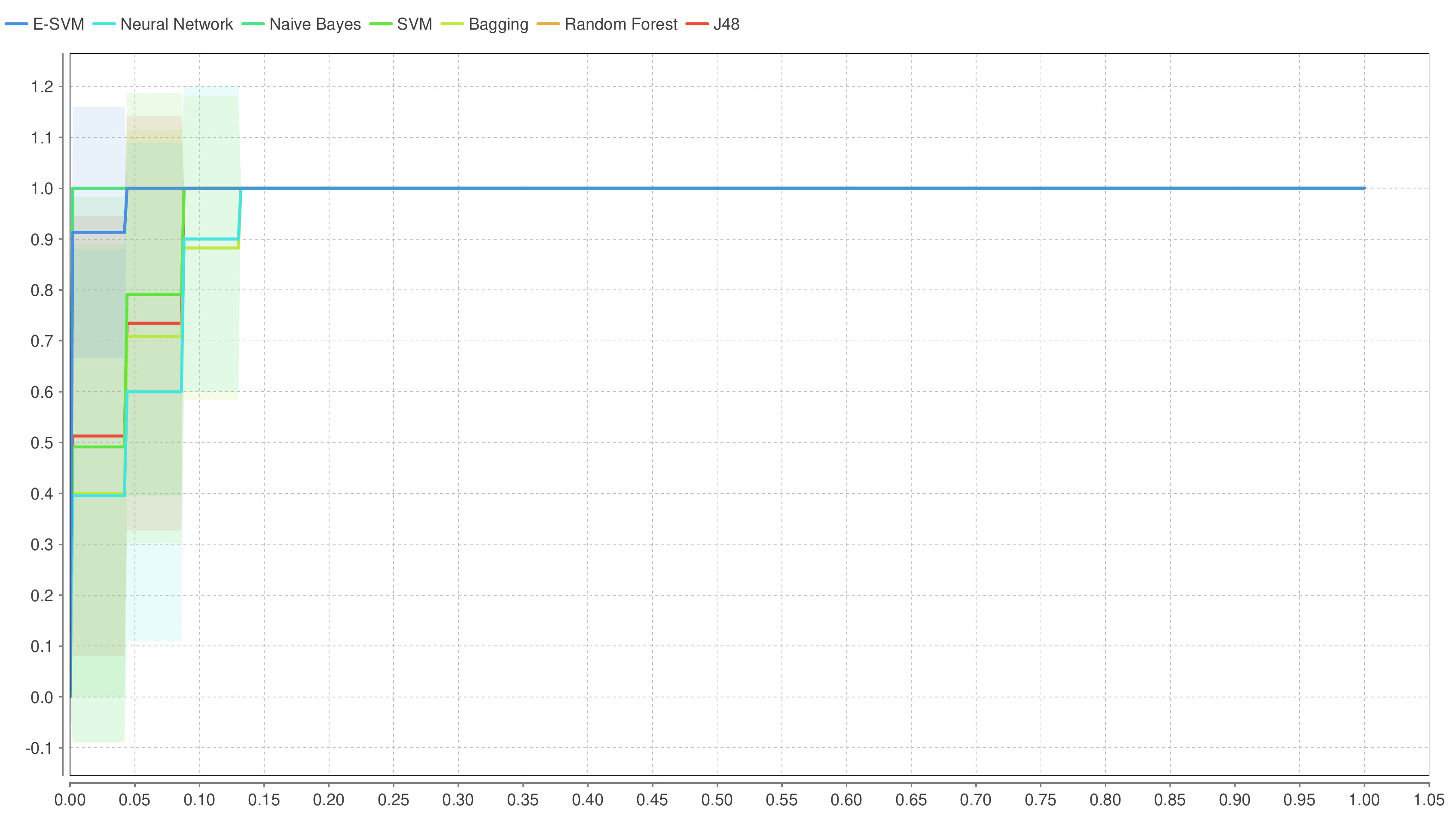}}
	\caption{ROC curve analysis for ML algorithms classification ability.}
	\label{fig7}
\end{figure}

ROC curve, which shows the classification ability of classifier (algorithms) is shown in Figure \ref{fig7}.
The ability of classifiers is predicted by a threshold value in ROC curve graphs. ROC is measured by the TP\footnote{ True Positive is use for XYZ} rate and FP\footnote{ False Positive for ABC}  rate. As shown in Figure \ref{fig7}, the performance of Naive Bayed and Random Forest is clearly much better as compared to other classifiers. In nutshell, Random Forest lead all the algoritms in XSS Attacj detection, we conclude that Random Forest is the suitable algorithm based on accuracy, f-measure, speed and ROC analysis.
\section{Conclusion}\label{sec-7}
In this research paper, we proposed a detection and prevention framework for XSS attacks in hybrid applications, where an attacker launches his attack on mobile resources through WebView and bypasses the sandbox mechanism using WebView APIs. To prevent these types of attacks, we propose our own detection system on Java object level, which have both detection and prevention functions through classification.

Our approach is based on ML algorithms. This article shows that ML algorithms approach to detect and prevent XSS attacks is quite efficient as compared to previous proposed systems, such as statistic approach for detection proposed in \cite{b7}. We took a variety of top machine learning classification algorithms such as Evolutionary E-SVM, Neural Network, Naive Bayes, SVM, Bagging, Random Forest, and  J48. To the best of our knowledge, there is no existing system which detects XSS attacks using ML algorithms. Moreover, to assess the classification capability of these different classifiers, we extracted new features and created our own dataset which is named as APK\_XSS\_ATTACKS, where we record 460 attacks. Based on detailed comparison, this article conclude that Random Forest outperformed all classifiers in term of accuracy and F-measure. 


\begin{thebibliography}{1}
	
	\bibitem{b1} Shashank Gupta, B.B.G. \emph{Cross Site Scripting (XSS) attacks and defense mechanisms: classification and state-of-the-art}.\hskip 1em plus
	0.5em minus 0.4em\relax International Journal of System Assurance Engineering and Management 2015, 8, 19.
	
	\bibitem{b2}	Security, W. WhiteHat \emph{Security Application Security Statistics Report}\hskip 1em plus
	0.5em minus 0.4em\relax WhiteHat Company: California CA, 2017; pp 1-60.
	
	\bibitem{b3}	Arjun Guha, S.K., Trevor Jim \emph{Using Static Analysis for Ajax Intrusion Detection}.\hskip 1em plus
	0.5em minus 0.4em\relax ACM 2009, 1-10.
	
	\bibitem{b4}	Bao, W.; Yao, W.; Zong, M.; Wang, D. \emph{Cross-site Scripting Attacks on Android Hybrid Applications}.\hskip 1em plus
	0.5em minus 0.4em\relax ICCSP '17 Proceedings of the 2017 International Conference on Cryptography, Security and Privacy 2017, 56-61.
	
	\bibitem{b5}	Jian Mao, M., IEEE, Jingdong Bian, Guangdong Bai, Ruilong Wang, Yue Chen, Yinhao Xiao, and Zhenkai Liang, Member, IEEE \emph{Detecting Malicious Behaviors in JavaScript Applications}.\hskip 1em plus
	0.5em minus 0.4em\relax IEEE Access 2018, 6, 12284-12294.
	
	\bibitem{b6}	Rita H Wouhaybi, D.S. \emph{HYBRID MOBILE INTERACTIONS FOR NATIVE APPS AND WEB APPS}.\hskip 1em plus
	0.5em minus 0.4em\relax Google 2017, 20.
	
	\bibitem{b7}	Jing Yu, T.Y. \emph{Access Control to Prevent Attacks Exploiting Vulnerabilities of WebView in Android OS. High Performance Computing and Communications}.\hskip 1em plus
	0.5em minus 0.4em\relax  IEEE International Conference on Embedded and Ubiquitous Computing (HPCC\_EUC), 2013 IEEE 10th International Conference on 2014, 1-6.
	
	\bibitem{b8}	Sungho Lee, J.D., Sukyoung Ryu \emph{HybriDroid: Static Analysis Framework for Android Hybrid Applications}.\hskip 1em plus
	0.5em minus 0.4em\relax ACM 2016, 1-12.
	
	\bibitem{b9}	Tongbo Luo, H.H., Wenliang Du, Yifei Wang, and Heng Yin \emph{Attacks on WebView in the Android System}.\hskip 1em plus
	0.5em minus 0.4em\relax ICPS: ACM International Conference Proceeding Series 2011, 343-352.
	
	\bibitem{b10}	William Enck; Machigar; Ongtang; Patrick, a.\emph{ McDaniel Understanding Android Security}.\hskip 1em plus
	0.5em minus 0.4em\relax IEEE Computer Society 2009.
	
	\bibitem{b11}	Xi Xiaoa, R.Y., Runguo Yeb, Qing Lia, Sancheng Pengc,Yong Jianga \emph{Detection and Prevention of Code Injection Attacks on HTML5-based Apps. Advanced Cloud and Big Data}.\hskip 1em plus
	0.5em minus 0.4em\relax  Third International Conference on 2015, 1-8.
	
	\bibitem{b12}	Jun Yang, P.Y., Xiaohui Jin, Qian Ma, \emph{ Multi-Classification for Malicious URL Based on Improved Semi-supervised Algorithm}.\hskip 1em plus
	0.5em minus 0.4em\relax International Conference on Computational Science and Engineering (CSE) and IEEE International Conference on Embedded and Ubiquitous Computing (EUC) 2017, 1-8.
	
	\bibitem{b13}	Nunan, A.E.; Souto, E.; dos Santos, E.M.; Feitosa, E., \emph{ Automatic classification of cross-site scripting in web pages using document-based and URL-based features}.\hskip 1em plus
	0.5em minus 0.4em\relax Computers and Communications (ISCC), 2012 IEEE Symposium on 2012, 000702-000707.
	
	\bibitem{b14}	Andreas Dewald, T.H., Felix C. Freiling, \emph{ ADSandbox: Sandboxing JavaScript to fight Malicious Websites}.\hskip 1em plus
	0.5em minus 0.4em\relax SAC '10 Proceedings of the 2010 ACM Symposium on Applied Computing 2010, 1859-1864.
	
	\bibitem{b15}	Thomas Blasing, L.B., Aubrey-Derrick Schmidt, Seyit Ahmet Camtepe, and Sahin Albayrak, \emph{ An Android Application Sandbox System for Suspicious Software Detection}. \hskip 1em plus
	0.5em minus 0.4em\relax Malicious and Unwanted Software (MALWARE), 2010 5th International Conference on 2010, 1-8.
	
	\bibitem{b16}	Michael Spreitzenbarth; Felix Freiling, F.E.; Thomas Schreck, J.H., \emph{ Mobile-Sandbox: Having a Deeper Look into Android Applications}. \hskip 1em plus  0.5em minus 0.4em\relax ACM 2013, 1-8.
	
	\bibitem{b17}	Lee, I.; Jeong, S.; Yeo, S.; Moon, J., \emph{  A novel method for SQL injection attack detection based on removing SQL query attribute values}.\hskip 1em plus
	0.5em minus 0.4em\relax Mathematical and Computer Modelling 2012, 55, 58-68.
	
	\bibitem{a17} Kosuga, Y., Kono, K., Hanaoka, M., Hishiyama, M., Takahama, Y. (2007, December). \emph{Sania: Syntactic and semantic analysis for automated testing against sql injection}. In Twenty-Third Annual Computer Security Applications Conference (ACSAC 2007) (pp. 107-117). IEEE.
	
	\bibitem{c17} Halfond, W. G., Orso, A. (2005, November). \emph{AMNESIA: analysis and monitoring for NEutralizing SQL-injection attacks}. In Proceedings of the 20th IEEE/ACM international Conference on Automated software engineering (pp. 174-183).
	
	\bibitem{b18}	Artzi, S.; Dolby, J.; Tip, F.; Pistoia, M., \emph{  Fault Localization for Dynamic Web Applications}.\hskip 1em plus
	0.5em minus 0.4em\relax IEEE Transactions on Software Engineering 2012, 38, 314-335.
	
	\bibitem{b19}	Johari2, S.S.a.R., \emph{  Survey of Cross-site Scripting Attack in Android Apps}.\hskip 1em plus
	0.5em minus 0.4em\relax 1-6.
	
	\bibitem{b20}	Iker Burguera and Urko Zurutuza, S.N.-T. Crowdroid, \emph{  Behavior-Based Malware Detection System for Android}.\hskip 1em plus
	0.5em minus 0.4em\relax SPSM '11 Proceedings of the 1st ACM workshop on Security and privacy in smartphones and mobile devices, 15-26 
	
	\bibitem{b21}	Jinyung Kim, Y.Y., and Kwangkeun Yi, Junbum Shin, \emph{SCANDAL: Static Analyzer for Detecting Privacy Leaks in Android Applications}.\hskip 1em plus
	0.5em minus 0.4em\relax 1-10.
	
	\bibitem{b22}	Herzberg2, A.D.B.B.a.M., \emph{  On the Static Analysis of Hybrid Mobile Apps}.\hskip 1em plus
	0.5em minus 0.4em\relax  A Report on the State of Apache Cordova Nation. 2016, 72-88.
	
	\bibitem{a22} Jin, X., Wang, L., Luo, T., Du, W. (2015). \emph{Fine-grained access control for html5-based mobile applications in android}. In Information Security (pp. 309-318). Springer, Cham.
	
	\bibitem{b23}	Korel, B., \emph{Automated software test data generation}.\hskip 1em plus
	0.5em minus 0.4em\relax IEEE Transactions on Software Engineering 1990, 16, 870-879.
	
	\bibitem{b24}	Andrea Avancini, M.C., \emph{  Security Testing of Web Applications: a Search Based Approach for Cross-Site Scripting Vulnerabilities}.\hskip 1em plus
	0.5em minus 0.4em\relax International Working Conference on Source Code Analysis and Manipulation 2011, 1-10.
	
	\bibitem{b25}	Moataz A. Ahmed, F.A., \emph{  Multiple-path testing for cross site scripting using genetic algorithms}.\hskip 1em plus
	0.5em minus 0.4em\relax Journal of Systems Architecture 2015, 64, 50-62.
	
	\bibitem{b26}	Elovici, A.S.Y.F.Y., \emph{ Automated Static Code Analysis for Classifying Android Applications Using  Machine Learning}.\hskip 1em plus
	0.5em minus 0.4em\relax International Conference on Computational Intelligence and Security 2010, 1-5 
	
	\bibitem{b27}	Ibéria Medeiros INESC-ID, N.N.L., Miguel Correia, \emph{INESC-ID Detecting and Removing Web Application Vulnerabilities with Static Analysis and Data Mining}.\hskip 1em plus
	0.5em minus 0.4em\relax IEEE Transactions on Reliability 2015, 65, 54-69.
	
	\bibitem{b28}	William Enck, P.G., Byung-Gon Chun,\emph{  TaintDroid: An Information-Flow Tracking System for Realtime Privacy Monitoring on Smartphones}.\hskip 1em plus
	0.5em minus 0.4em\relax ACM Transactions on Computer Systems (TOCS) 2014, 32, 1 - 15.
	
	\bibitem{b29}	CARSTEN; WILLEMS; THORSTEN; HOLZ , F.F., Florian Echtler, \emph{  Toward Automated Dynamic Malware Analysis Using CWSandbox}.\hskip 1em plus
	0.5em minus 0.4em\relax IEEE Computer Society 2007, 5, 1 - 8.
	
	\bibitem{b30}	Shashank Gupta, B.B.G.,\emph{  CSSXC: Context-Sensitive Sanitization Framework for Web Applications against XSS Vulnerabilities in Cloud Environments}.\hskip 1em plus
	0.5em minus 0.4em\relax Procedia Computer Science 2016, 85, 198-205.
	
	\bibitem{b31}	Abdalla Wasef Marashdih, Z.F.Z.H.K.O.,\emph{  Web Security: Detection of Cross Site Scripting in PHP Web Application using Genetic Algorithm}.\hskip 1em plus
	0.5em minus 0.4em\relax (IJACSA) International Journal of Advanced Computer Science and Applications, 2017, 8, 1 - 12.
	
	\bibitem{b32}	Fabien Duchene, R.G., Sanjay Rawat, Jean-Luc Richier, \emph{ XSS Vulnerability Detection Using Model Inference Assisted Evolutionary Fuzzing}.\hskip 1em plus
	0.5em minus 0.4em\relax Fifth International Conference on Software Testing, Verification and Validation 2012, 1 - 3.
	
	\bibitem{b33}	Rapidminer, \emph{  Data Science Platform for Analytics}.\hskip 1em plus
	0.5em minus 0.4em\relax Available online: https://rapidminer.com/ 
	
	\bibitem{IEEEhowto:kopka}
	H.~Kopka and P.~W. Daly, \emph{A Guide to \LaTeX}, 3rd~ed.\hskip 1em plus
	0.5em minus 0.4em\relax Harlow, England: Addison-Wesley, 1999.
\end{thebibliography}
\end{document}